\documentstyle[twoside,fleqn,espcrc2,epsf]{article}

\newcommand{\AmS}{{\protect\the\textfont2
  A\kern-.1667em\lower.5ex\hbox{M}\kern-.125emS}}

\title{Real time correlations at finite Temperature for the Ising model.}

\author{E. Mendel \thanks{Presented at the Lattice98 Conference at Boulder, Co.}\\
 { FB Physik,
 Carl von Ossietzky Universit\"at Oldenburg,
         26111 Oldenburg, Germany}}      
\begin{document}

\begin{abstract}
  After having developed a method that measures real time
evolution of quantum systems at a finite temperature, we
present here the simplest field theory where this scheme
can be applied to, namely the $1+1$ Ising model.
  We will compute the probability that if a given spin is
up, some other spin will be up after a time $t$, the
whole system being at temperature $T$. We can thus study
spatial correlations and relaxation times at finite $T$.
  The fixed points that enable the continuum real time limit
can be easily found for this model. 
  The ultimate aim is to get to understand real time evolution
in more complicated field theories, with quantum
effects such as tunneling at finite temperature.       
\end{abstract}

\maketitle

\section{Introduction}

  The description of processes in real time for quantum systems at finite
temperature is a hard subject that has, to my knowledge, not been fully
understood. Most work on this field is based on the linear response
approximation, that leads to computation of real time correlation
{\em amplitudes} of operators in a ``thermal state'' mixture.
We have developed a method in which one computes directly transition
{\em probabilities} among observables for subsets of the whole quantum
system. If one wants to treat consistently a system at finite temperature
but still measure some time correlations, this can only make sense for
small subsystems of a large system in thermal equilibrium. In the
thermodynamic limit these correlations should converge to an answer,
making the use of an external bath consistent with the full quantum
treatment. We will see that this formalism gives a good probabilistic
interpretation, in contrast to the one obtained from the usual treatment.

  In an earlier work \cite{MenNes} we have presented the formalism 
for several problems in quantum mechanics with one or a few degrees
of freedom, showing interesting phenomena such as real time tunneling at
finite $T$, but still without being able to show the self consistency 
expected in the thermodynamic limit mentioned above. For this purpose
and to show that the same formalism can be applied to field theories,
the $1+1$ dimensional Ising model will be computed in this work. In this
model, also called the quantum Ising model, the spins in the spatial
dimension will be kept at discrete distances while the time axis
(real or Euclidean) will reach the continuum. In this process the
coupling constants have to be renormalized so as to get a proper continuum
limit. These renormalization trajectories were known
for the Euclidean Hamiltonian formalism \cite{SusFra} and will be
obtained by analytical continuation for the real time case. The lack
of such analytical expression for more complicated theories, such as
for the interesting electroweak one to study Baryon Number generation \cite{Smit},
could turn out to be a big hurdle. The gauge fields can be approximated
by discrete subgroups to get finite dimensional transfer matrices as for
the spin system.
\section{Spin correlations in time}
  To fix ideas let us ask for the probability, with the quantum Ising model,
for some spin to be up at time $t$ if a measured one was up at $t=0$, the
whole system being at finite $T$.

 In order to compute such a correlation, we just know 
that the system was in a {\em mixed state} described by 
$ \rho = \exp(-\beta H) $.
Then we measure at $t=0$ a spin  $s_1$ (without
loss of generality we measure the first spin in a closed ring, the others staying undetermined),
corresponding to applying the projector (with a given $s_1=\pm 1$):       
\begin{equation}
 P_{s_1} = |s_1 \rangle \langle s_1| * {\boldmath \rm I \ }_{ 2,..N} 
\end{equation}
on {\em both sides} of $\rho$.  
This operator evolves in $t$ as usual, 
$ U^+_t ( P_{s_1} \rho P_{s_1} ) U_t$ , with $ U_t = \exp(-i H t)$ .

We measure again at a time $t$ some spin $S_i$ with the corresponding 
projectors $ P_{S_i}$.  The probability is then:  
\begin{equation}
   P_t^{\beta} (s_1,S_i) = \frac{1}{\rm Norm.} Tr[ P_{S_i} U^+_t ( P_{s_1} \rho P_{s_1} ) U_t ]    
\end{equation}
where we have discarded one of the  $P_{S_i}$  due to the cyclicity of the trace.
The Normalization can be picked so that the probability for every $S_i$ to be up or down adds to $1$.
 This expression can be seen as a proper transition probability in full analogy to the quantum 
mechanics case \cite{MenNes} by going to the energy representation. Notice that we need three
projectors inserted for this to happen, if one could commute $ [\rho, P_{s_1}]$ one would 
get the more usual correlations \cite{Smit} which do not have a clear probability interpretation. In the classical
limit both expressions would coincide.

  Equation (2) can be recast in a suitable form for using the path integral formalism     
$$
P_t^\beta(s_1,S_i)=\sum_{s_{j=2..n}}^\pm   \sum_{s'_{k=2..n}}^\pm \sum_{S_{l \ne i}}^\pm 
   \langle s_1..s_j..| \rho |s_1..s'_k..\rangle 
$$             
\begin{equation}
\cdot \langle s_1..s'_k..|U^+_t|..S_i..S_l..\rangle \langle ..S_i..S_l..|U_t |s_1..s_j..\rangle,
\end{equation}
obtaining a product of three Green functions.
Each of these Green functions can be calculated 
by multiplying the transfer matrix for infinitesimal "$dt$" by
itself N times to obtain times $ t=(2^N)*dt$. This method works also for 
real $t$ as we are considering all the contributing phases.

  The transfer matrix that takes us from one slice (with spins $s_i$)
to the next (spins $s'_j$), has $2^n \times 2^n$ elements for the usual Ising model:
\begin{equation}
M({\b s} ; {\b s}')=\! \prod_{i=1..n} \! e^{\frac{1}{2} K_s (s_i s_{i+1} + s'_i s'_{i+1}) + \frac{1}{2} K_\tau (s_i - s'_i)^2}
\end{equation}             
  In order to consider this to be the short time propagator of a quantum system, one could naively reinterpret
the couplings $(K_s, K_\tau)$ as new ones $(\Delta \tau K'_s, K'_\tau/\Delta \tau)$, for which one can easily see
that in Eq. (4) the exponent looks like an action with potential and kinetic energy terms, as in  first quantization.
The problem that one encounters in this naive approach is that this transfer matrix does not scale correctly for
decreasing $\Delta \tau$, as we need to reach criticality for continuum $ \tau$. Luckily for the asymmetric Ising
model in 2 dimensions one can use duality to find a critical line  at: $\sinh(K_s)\sinh( K_\tau)=1$. This was used
to get scaling in the continuum  Euclidean time  case \cite{SusFra}, by taking $K_s(\Delta \tau)=K'_s \Delta \tau$ and then
$K_\tau =\frac{1}{2}{\rm arcsinh}(1/\sinh(2 \Delta \tau))$. 
 The new ingredient comes in when one wants to define the transfer matrix for real times $\Delta t$. I assumed
analytical behavior in the phase transition line, so that for $K_s(\Delta t)= i \ K'_s \Delta t$ we would get criticality with
\begin{equation}
K_t =\frac{1}{2}\ {\rm arccosh}(1/\sin(2 \Delta t))\ - \ i \pi /4 
\end{equation}             
obtaining a complex coupling $K_t$ with a small but important phase (for scaling).
With these couplings for the real time propagators we have been able to get very good convergence to the
continuum limit with a few steps of multiplying the transfer matrix by itself. The convergence is easy to test
by using the composition rule for Green's functions, $M_{2\Delta t}=M_{\Delta t}\cdot M_{\Delta t}$. We still have
the freedom to dial $K'_s$ in order to have a stronger (weaker or even antiferro.) spatial binding. 

  The three Green's functions needed in Eq.(3) can then be computed by multiplying  the Euclidean or real time
transfer matrices with the couplings as seen above. One needs $N$ matrix multiplications by itself to reach
times $\beta=2^N \Delta \tau$ or $t=2^{N'} \Delta t$. In order to construct numerically the transfer matrix 
it is very convenient to define first an array S($2^n$, $n$) giving for each possible spin configuration in space,
the value of every spin. With this array Eq.(4) is easily rewritten and $M_{\Delta t}({\b s} ; {\b s}')$ can be computed
and stored for each pair of spin configurations ${\b s}$ and  ${\b s}'$. After multiplying the $M$'s by themselves
to obtain the Green's functions one still has to compute the product of the 3 Green's functions, making sure that
the observables $s_1$ and $S_i$ (for a given $i$) have a fixed value and all the other ones get summed up. With
this matrix method it is easy to get up to 12 or 14 spins on a spatial ring, with no stringent limits on $t$ or $\beta=1/T$.

\begin{figure}[thb]
\vspace{-1mm}
\hspace{-9mm}
\epsfxsize=83mm
\epsffile{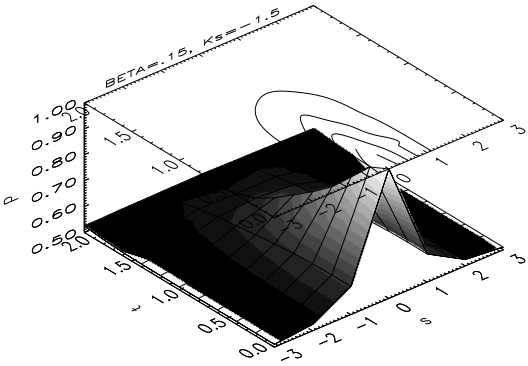}
\vspace{-12mm}
\caption{Probability vs. space and time to find the spins aligned at high $T$. } 
\end{figure}

\begin{figure}[hbt]
\vspace{-5mm}
\hspace{-9mm}
\epsfxsize=83mm
\epsffile{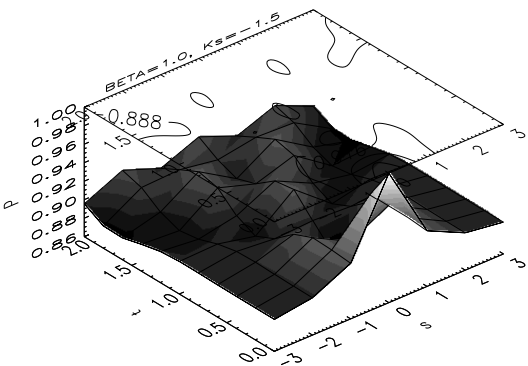}
\vspace{-12mm}
\caption{Prob. for  spin alignment at lower $T$.} 
\vspace{-2mm}
\end{figure}

  We have made various simulations in order to measure the probabilities, $P_t^\beta(s_1,S_i)$, to get a spin up at position
$i$ after a time $t$ of having prepared  $s_1$ up. At high Temperatures the prepared spin is initially aligned with its 
neighbors with a short spatial correlation length, the whole signal relaxing in time to eventually disappear ($P=50\%$ 
everywhere) as seen in fig.1. At intermediate $T$'s the initial spatial correlation length  and the relaxation time are larger,
as the fluctuating spins cannot screen the signal as at high $T$'s. At even lower $T$'s the spins get all almost aligned
initially and continue in this situation for long times, as if going into an ordered phase, with interesting oscillating
spin waves on top of the alignment, as seen in fig.2.

 By increasing the coupling  $K'_s$, the stiffness grows, as if we had effectively decreased the temperature, and in fact
we get similar phenomena  as above but at already higher $T$'s. At very weak $K'_s$ the prepared spin is almost uncoupled
with the rest and mainly just oscillates down and up again. Changing the sign of $K'_s$ we even get antiferromagnetic behavior
among the spins.

\vspace{-1mm}
\section{Conclusions}

  We have worked out for the $1+1$ dimensional quantum Ising model the spin probability correlations in real time
and finite $T$. We find results that seem very plausible if one could design an experiment to measure these
correlations. Besides the interesting results for this model, this work  shows that this method can be applied to
field theories as well as for simple quantum systems. Moreover we have been able to check for this model that
if we increase the number of spins on a (larger) spatial ring, we reach limiting functions for the correlations,
thus reaching the thermodynamic limit. In this limit the whole scheme is consistent in the sense that the initial
preparation of spin $s_1$ does not take the full system out of thermodynamic equilibrium and the rest of the
system can be considered as a genuine quantum heat bath to study the time evolution.    

 Our ultimate goal is to find a way to treat the field theory case of tunneling
with instantons in real time and finite $T$. As we have seen, it is in principle
doable with these methods but will be hard to implement due
to the number of degrees of freedom after discretizing the gauge fields. The other hard part will consist
in finding the renormalization trajectories for the coupling, which could lie in the compex plane, in order
to reach a sensical continuum time limit.

 I would like to thank J. Polonyi, P. Rujan and  L. Polley for useful discussions on this work.
 

\begin{thebibliography}{00100}
\bibitem{MenNes} E. Mendel and M. Nest, Nucl. Phys. B 63 (Proc. Suppl) (1998) 445;
extended version in Preprint hep-th/9807030. 
\bibitem{SusFra} E. Fradkin and L. Susskind, Phys. Rev. D 17 (1978) 2637. 
\bibitem{Smit} J. Smit, Nucl. Phys. B 63 (Proc. Suppl.) (1998) 89 and
refs. therein.
\end{thebibliography}
\end{document}